\documentclass[english,reprint,showkeys]{revtex4-1}
\usepackage[T1]{fontenc}
\usepackage[latin9]{inputenc}
\usepackage{amstext}
\usepackage{graphicx}

\makeatletter

\providecommand{\tabularnewline}{\\}

\makeatother

\usepackage{babel}
\begin{document}

\title{The COVID-19 pandemic: growth patterns, power law scaling, and saturation}

\author{H.M. Singer}

\date{2020-04-07}

\affiliation{Institute for Complex Systems Research ICSR\break 8820 Wädenswil,
Switzerland}
\email{hsinger@icsr.ch}

\selectlanguage{english}%

\begin{abstract}
More and more countries show a significant slowdown in the number
of new COVID-19 infections due to effective governmentally instituted
lockdown and social distancing measures. We have analyzed the growth
behavior of the top 25 most affected countries by means of a local
slope analysis and found three distinct patterns that individual countries
follow depending on the strictness of the lockdown protocols: exponential
rise and fall, power law, or logistic. For countries showing power
law growth we have determined the scaling exponents. For countries
that showed a strong slowdown in the rate of infections we have extrapolated
the expected saturation of the number of infections and the expected
final date. Two different extrapolation methods (logistic and parabolic)
were used. Both methods agree on the order of magnitude of saturation
and end dates. Global infections rates are analyzed with the same
methods. The relevance and accuracy of these extrapolations is discussed.
\end{abstract}
\maketitle

\section*{Introduction}

In December 2019 a novel corona virus (COVID-19) was discovered in
Wuhan, China. The infections quickly spread first within China but
soon thereafter moved to neighboring Asian countries and as of February
2020 reached all continents. The first massive outbreak outside mainland
China happened in Northern Italy from which it spread rapidly all
over Europe. As of April 5th 23:00 CET over 1'270'000 people in 184
countries and regions have been infected \cite{arcgis2020_04_05}.
Currently, the most affected country is the USA with over 337310 infections.

Due to its high infectivity and since no medical cure or vaccine is
available yet governments around the world have instituted drastic
measures such as curfews, lockdowns and social distancing campaigns
to slow down the number of new infections and reduce the rate at which
the virus spreads. 

The mortality rate has been found to be 2.5\%\cite{LAI2020105924}
(comparison: SARS 9.6\% \cite{SMITH20063113} and H1N1 influenza 0.6\%\cite{VaillantEtAl2009})
but the severity of cases is much higher for the elderly population
\cite{jcm9020523} and people with preexisting health conditions like
cardiovascular and respiratory diseases, diabetes, or cancer \cite{DRIGGIN2020,jcm9020523}
as well as heavy smoking \cite{Vardavas2020} where the mortality
rates can increase to 5-11\% \cite{jcm9020523}.

Increasingly more countries affected by the COVID-19 virus find a
reduction in the rate of new infections due to strict lockdowns. While
no country so far has reached the same saturation of the total number
of infections as China many seem to be on the way towards this stage
showing a similar situation as China at the end of February 2020. 

Traditionally biological systems such as the spreading of a disease
are modeled as exponential growth leading to the differential equation
\[
\frac{dN}{dt}=\lambda N
\]
where $N(t)$ is the cumulative number of infections and $\lambda>0$
is the growth rate. In the case of an unlimited supply of susceptible
individuals it follows immediately that $N(t)=e^{\lambda t}$. Of
course, in any real system, the number of susceptible individuals
in finite. Hence, as the number of infections grows the rate will
slow down and eventually the total number of infections will saturate.
This can be modeled as
\begin{equation}
\frac{dN}{dt}=\lambda N\left(1-\frac{N}{K}\right)\label{eq:logeq}
\end{equation}
with $K>0$ the carrying capacity. The solution to eq. (\ref{eq:logeq})
is the Verhulst logistic curve \cite{Verhulst1838}
\begin{equation}
N(t)=\frac{K}{1+Be^{-rt}}\label{eq:logcurve}
\end{equation}
with 
\[
B=\frac{K-N_{0}}{N_{0}}
\]
and $N_{0}$ the initial infected population size and $N_{\infty}=\lim_{t\rightarrow\infty}N(t)=K$.

The logistic model has been used to model the COVID-19 pandemic by
several research groups \cite{jia2020prediction,yang2020rational,batista2020.02.16.20023606,wu2020generalized,vattay2020predicting,dattoli2020note,dattoli2020evolution,morais2020logistic,kumar2020epidemiological,sonnino2020dynamics,ttrai2020covid19}.

The goodness of fit for the logistic model and other models were studied
for the China data in \cite{jia2020prediction,yang2020rational,wu2020generalized}.
It was found that for China the logistic model underestimates the
final number of infections \cite{yang2020rational} and also that
it does not predict the correct growth rates while still in the strong
growth regime \cite{sonnino2020dynamics}. A quality criterion to
determine the goodness of expected number of infections was determined
in \cite{ttrai2020covid19}. Data from Italy under the assumption
of logistic growth was studied in \cite{vattay2020predicting,sonnino2020dynamics,morais2020logistic,dattoli2020note,dattoli2020evolution}. 

The problem with the logistic model is that it assumes an exponential
growth which is successively clamped off by the diminishing reservoir
of people to be infected. Data analysis for the China data \cite{Ziff2020.02.16.20023820,li2020scaling}
and for the top 25 affected countries \cite{singer2020shortterm}
as well as additional studies for particular countries \cite{manchein2020strong,blasius2020powerlaw}
have clearly shown, however, that by the introduction of lockdown
measures and social distancing the growth phase is not exponential
any more but follows a power law. The scaling exponents are country-specific
and depend on the strictness of governmental interventions and the
actual adherence of the people. The power law growth is connected
with the actual societal structure that has been shown to exhibit
small world or scale-free \cite{Watts1998,Watts2004} behavior. The
spread of disease is determined by the number of human interactions
and therefore limiting those will control the rate at which an infection
can spread. After a short initial exponential growth period the institution
of a lockdown reduces the possible spreading avenues: the number of
infections as well as the number of recovered and deceased patients
follows a power law
\[
N(t)=Bt^{\gamma}
\]
until a slow down of the rate of infection in noticeable due to a
lack of susceptible hosts.

The reason why a power law growth is observed is as follows: Assuming
a densely populated area like a major city or town each individual
will during the course of a day have many different interactions with
other people. Some of them will be with close relations, i.e. people
that are important for the individual such as spouse, children, family,
friends, or coworkers. Others, however will be random contacts when
sharing the same space for example in public transport, shops, entertainment
facilities etc. For the spreading of a disease the importance of the
contact, whether or not it is close or random, is not relevant. The
important thing, from the point of view of a virus, is close physical
proximity. This might be given at the dinner table, in the office,
riding a bus or train, or walking in the public park. In fact, on
an average day the number of random contacts that satisfy the transmission
criterion might well be much larger than the contacts an individual
has with his close relations. Random contacts however are the central
ingredient for an exponential growth. By instituting lockdown measures,
social distancing and telling people to stay at home it is essentially
possible \textendash{} if properly followed \textendash{} to eliminate
all random contacts leaving only the close relations of family and
work. It has been observed that the networks spanned by those relationships
follow a power law behavior \cite{Watts1998,Watts2004}. Since the
virus has now only this network as means of spreading it follows that
the number of infections will also show a power law dependence. In
reality it is, of course, not possible to eliminate all random contacts
(for example interactions in grocery shops), but so far the data shows
that the stricter the lockdown measures are and the more people adhere
to it the closer to a power low the rate of infections behaves.

\section*{Method}

All data related to COVID-19 was downloaded from the publicly available
JHU-CSSE (2020) data source provided continuously by the Johns Hopkins
University Center for Systems Science and Engineering (JHU CSSE) \cite{csse_github_2020_04_05}.
Aggregate data calculation for countries with different regions or
states were performed after collecting by a simple summation over
all associated regions or provinces (for the countries China, Australia,
and Canada). 

We have selected the data sets with the 25 most infections (and China)
as presented in \cite{arcgis2020_04_05} on 2020-04-05 23:00 CET.
The countries and their infection counts are given in Table \ref{tab:Countries-infections}.

In our previous publication \cite{singer2020shortterm} we have determined
the scaling coefficients $\gamma$ for each of the 25 most affected
countries and compared the relative growth rates with each other.
Since more data is available by now we have analyzed the local slopes
of the presumed power law range in further detail. In a log-log representation
$\log N(t)$ vs. $\log t$ a power law $N(t)\sim Bt^{\gamma}$ becomes
\[
\log N(t)=\gamma\log(x)+\log(B)
\]
Hence, the derivative is
\[
\frac{d\log N(t)}{d\log(t)}=\gamma=\text{const.}
\]
We have calculated the discrete derivative (local slope)
\[
\frac{\Delta\log N(t)}{\Delta\log t}
\]
over the available range and plotted it against the time to analyze
the power law behavior over time. A power law behavior would exhibit
regions of constant values, whereas monotonically increasing values
would indicate exponential growth and decreasing values exponential
slow down. This plot serves as an amplifier of the internal structures
governing the growth rate to unveil changes that cannot be seen in
a simple log-log representation of the data.

\begin{figure}
\includegraphics[scale=0.6]{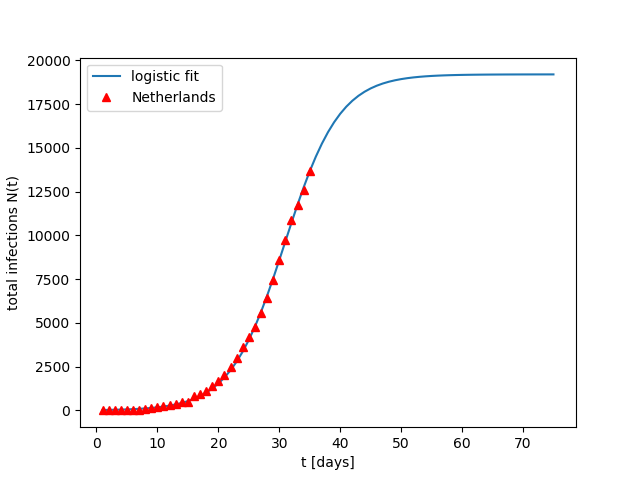}

\caption{\label{fig:Logistic-fit}Logistic fit for the total number of infections
in the Netherlands. }

\end{figure}

On data sets where we have found a significant downward trend of this
slope (i.e. a slowdown of the growth rate) we have also performed
two extrapolation methods in order to determine the presumed end of
the infection in that particular country. In Fig. \ref{fig:Logistic-fit}
the infection data for the Netherlands is plotted with a logistic
equation fit.

\begin{figure}
\includegraphics[scale=0.6]{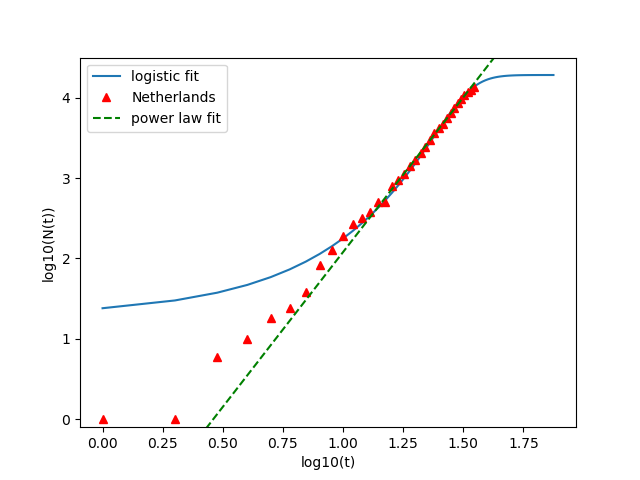}

\caption{\label{fig:Log-log-representation logistic fit}Log-log representation
of the total number of infections in the Netherlands with superimposed
power law and logistic fit. Clearly the power law fits the data much
better than the logistic equation, particularly in the beginning of
the infection spread.}
\end{figure}

At first the glance the logistic equation fits the data well. However
a closer examination in the log-log representation shows that the
logistic equation fits the data only well closer to the saturation
point when the growth rate is already substantially slowing down.
The log-log representation of the total number of infections in the
Netherlands together with the superimposed power law fit and the logistic
fit are given in Fig. \ref{fig:Log-log-representation logistic fit}.
It can be seen clearly that the data follows a power law behavior
in most of the available data range and that the logistic fit is particularly
bad in the beginning of the growth phase.

\begin{figure}
\includegraphics[scale=0.6]{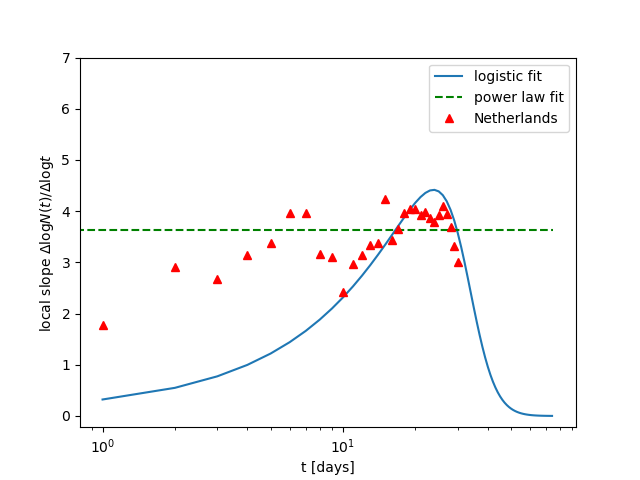}

\caption{\label{fig:Local-slope-calculation}Local slope calculation for the
number of infections in the Netherlands. Power law fit and logistic
fit are superimposed. Clearly the logistic fit cannot accurately describe
the data. The data seems to follow a power law over a long range of
the available data.}
\end{figure}

The local slope calculation is shown in Fig. \ref{fig:Local-slope-calculation}
for the infection data in the Netherlands. The best fit for the logistic
equation and the power law fit is superimposed. As can be seen clearly
the local slope remains constant over nearly the whole data range
and only the last three data points show a substantial decrease in
the local slope and a close following of the logistic equation towards
the saturation. It is also obvious that the local slopes do not correspond
to what one would expect from a logistic curve. In the case of China
it was observed that the closer to the saturation the data tended
the better a fit to the logistic equation was found \cite{jia2020prediction,yang2020rational,wu2020generalized,batista2020.02.16.20023606}.

The logistic extrapolation performs a non-linear fit to equation (\ref{eq:logcurve})
and determines the parameters $K$, $B$, and $r$ by a non-linear
least square calculation (Levenberg- Marquardt \cite{levenberg1944,marquardt1963}).
Since by definition the final number of predicted infections is never
reached we have chosen the time $t$ at which 95\% of the final value
is reached to be indicative for the end date. Therefore the extrapolated
values for $N_{\infty,\text{log}}$ and $t|_{N(t)=0.95N_{\infty,\text{log}}}$are
calculated.

The parabolic fit \cite{li2020scaling} calculates a least square
fit to an inverted parabola in the log-log plot $y=ax^{2}+bx+c$.
The extremal point $x_{0}$ determines the time when the full saturation
has been reached and $y_{0}$ the total number of infections by:
\[
x_{0}=\frac{-b}{2a},\,y_{0}=ax_{0}^{2}+bx_{0}+c
\]
and

\[
N_{\infty,\text{par}}=10^{y_{0}},\,t_{\text{par}}=10^{x_{0}}
\]

An example of this extrapolation is given in Fig. 4 for the total
number of infections in Switzerland. The extrapolated extremum is
marked with an arrow.

\begin{figure}
\includegraphics[scale=0.6]{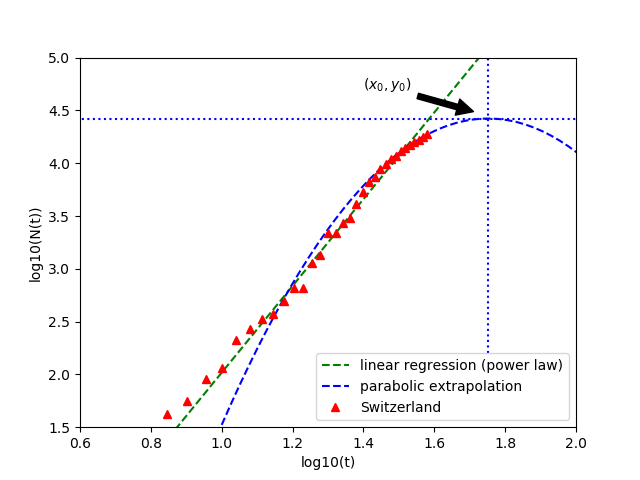}

\caption{\label{fig:Parabolic-extrapolation}Parabolic extrapolation of the
total number of infections in Switzerland. From the extremum point
$(x_{0},y_{0})$ the extrapolated final number of infections $N_{\infty,\text{par}}$
and the date $t_{\text{par}}$ at which this number is reached can
be calculated.}
\end{figure}

In order to obtain a more accurate estimation we have performed the
extrapolation calculations starting with the last 5 data points and
extending the number of data points backwards up to the last 30 data
points. If the predicted parameters $t_{\text{par}}$ and $N_{\infty,\text{par}}$
for the parabola fit and $t_{\text{log}}|_{N(t)=0.95N_{\infty,\text{log}}}$and
$N_{\infty,\text{log}}$ for the logistic fit were plausible, that
is $t_{\text{prediction}}>t_{\text{current}}$ and $N_{\infty}>N(t_{\text{current}})$,
then they were aggregated to an average value $\overline{t}_{\text{log}}$
and $\overline{t}_{\text{par}}$ and $\overline{N}_{\infty,\text{log}}$
and $\overline{N}_{\infty,\text{par}}$ respectively including their
standard deviations.

It should be noted that the right part of the inverted parabola after
the extremum $(x_{0},y_{0})$ does not have any physical significance.

Additionally the parabola fit is very sensitive to the data. If the
data points show a temporary increase again the parabola might flip
sides and will find the extremum on the left of the data with an $x_{0}$
that is smaller than the most recent data point or might flip such
that it points upwards. In such cases an extrapolation was not possible.
With respect to local deviations of the data towards the saturation
the non-linear logistic fit is much more stable and will find reasonable
fits also in situations where the parabolic fit fails.

\section*{Results}

We have analyzed the local slope of all the countries given in Table
\ref{tab:Countries-infections}. As indicated we have additionally
performed extrapolations for the countries which showed a significant
slowdown in their local slope plot. As of 2020-04-05 apart from China
20 of the top 25 countries showed signs of a slowdown in the rate
of infections, the 5 remaining countries are still in a strong growth
phase. We have calculated the scaling exponent $\gamma$ for these
countries and summarized them in Table \ref{tab:Countries-exponents}.

\begin{table}
\caption{\label{tab:Countries-infections}Countries with the most COVID-19
infections (2020-04-05)}
\begin{tabular}{|l|c|c|} 
\hline Country & \#Infections (2020-04-05) & Slowdown \\ \hline \hline 
World  &  1272115  &   \\\hline 
US  &  337072  &  y \\\hline 
Spain  &  131646  &  y \\\hline 
Italy  &  128948  &  y \\\hline 
Germany  &  100123  &  y \\\hline 
France  &  92839  &  y \\\hline 
China  &  82602  &  y \\\hline 
Iran  &  58226  &  y \\\hline 
United Kingdom  &  47806  &  y \\\hline 
Turkey  &  27069  &  n \\\hline 
Switzerland  &  21100  &  y \\\hline 
Belgium  &  19691  &  y \\\hline 
Netherlands  &  17851  &  y \\\hline 
Canada  &  15746  &  n \\\hline 
Austria  &  12051  &  y \\\hline 
Portugal  &  11278  &  n \\\hline 
Brazil  &  11130  &  y \\\hline 
South Korea  &  10237  &  n \\\hline 
Israel  &  8430  &  y \\\hline 
Sweden  &  6830  &  n \\\hline 
Norway  &  5687  &  y \\\hline 
Australia  &  5687  &  y \\\hline 
Russia  &  5389  &  y \\\hline 
Ireland  &  4994  &  y \\\hline 
Czech Republic  &  4587  &  y \\\hline 
Chile  &  4471  &   n\\\hline 
Denmark  &  4369  &  n \\\hline
\end{tabular}
\end{table}

\begin{table}
\caption{\label{tab:Countries-exponents}Countries with power law growth that
do not show sign of a slowdown. Calculation of the current scaling
exponent.}
\begin{tabular}{|c|c|}
\hline 
Country & $\text{\ensuremath{\gamma}}$\tabularnewline
\hline 
\hline 
Turkey & 4.90\tabularnewline
\hline 
South Korea  & 0.72\tabularnewline
\hline 
Portugal & 4.61\tabularnewline
\hline 
Sweden & 5.17\tabularnewline
\hline 
Denmark & 1.64\tabularnewline
\hline 
\end{tabular}
\end{table}

\begin{figure}
\includegraphics[scale=0.6]{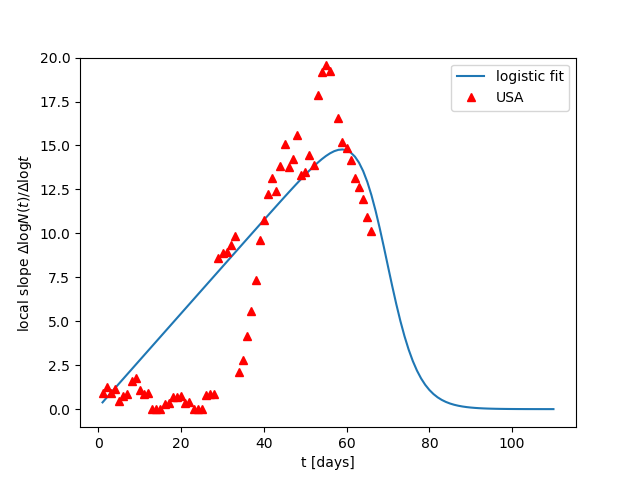}

\caption{\label{fig:Local-slope-plotUS}Local slope plot of US. A long period
of no cases followed by an exponential increase and a sharp downturn
soon thereafter. The logistic fit does not capture the behavior accurately.}
\end{figure}

\begin{figure}
\includegraphics[scale=0.6]{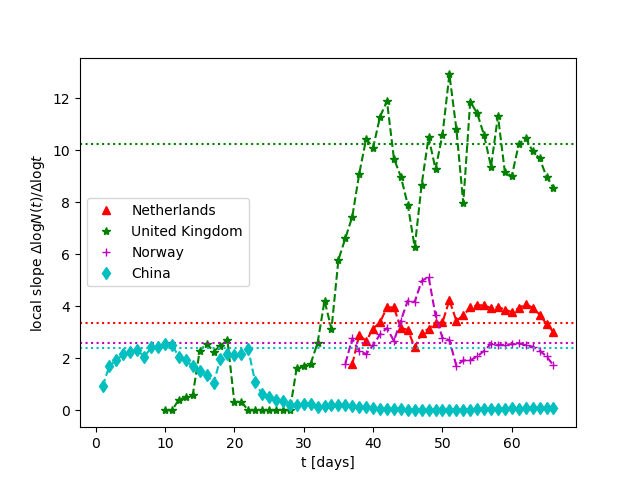}

\caption{\label{fig:Local-slope-plotPowerlaw}Local slope plot of countries
exhibiting power law growth. }
\end{figure}

\begin{figure}
\includegraphics[scale=0.6]{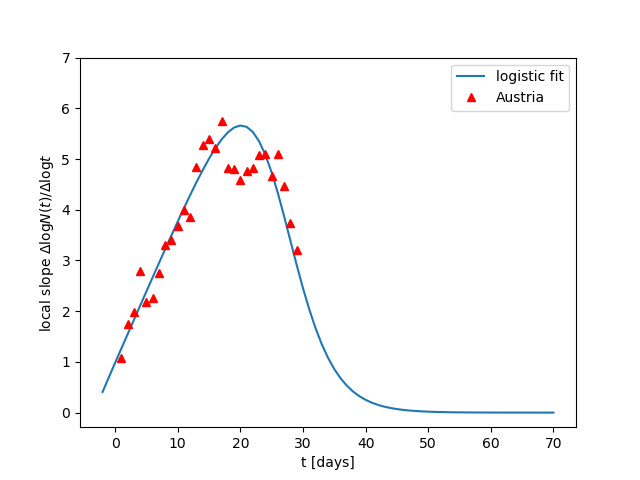}

\caption{\label{fig:Local-slope-plotAustria}Local slope plot of Austria. The
local slope follows the logistic equation.}
\end{figure}

While investigating the local slope of different countries we have
noticed that not all follow them same growth behavior. We have identified
3 different growth patterns that show very different behaviors:
\begin{enumerate}
\item \textbf{rise and fall pattern (exponential):} the total number of
infections stays very low for a long time and then suddenly shoots
up dramatically and exponentially. After a short period of time it
slows down equally rapidly with the rate of the slow down being exponential
as well. Example countries are US, France, Canada, and Australia.
A plot of such a behavior is shown in Fig. \ref{fig:Local-slope-plotUS}
for the USA. 
\item \textbf{power law pattern:} after an initial increase the growth becomes
a power law and stays for most of the time there. Only at the end
of the epidemic it slows down again and can be fitted by a logistic
equation. Examples of countries with this behavior are China, the
Netherlands, Norway, Czech Republic, Chile, and Russia. A plot of
the power law behavior is given in Fig. \ref{fig:Local-slope-plotPowerlaw}
for a selection of different countries.
\item \textbf{logistic pattern:} the growth follows quite closely the shape
of a logistic equation. Examples of countries are Austria, Switzerland,
Ireland, and Israel. A plot of such a behavior is shown in Fig \ref{fig:Local-slope-plotAustria}
for Austria.
\end{enumerate}

\begin{figure}
\includegraphics[scale=0.6]{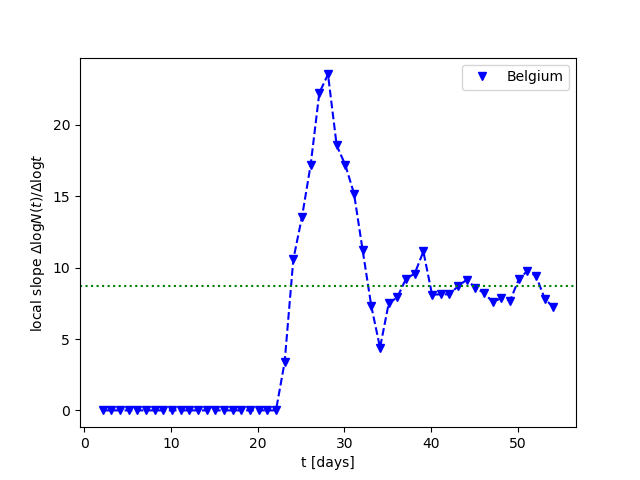}

\caption{\label{fig:Local-slope-Belgium}Local slope plot of Belgium. At first
the data displays the typical rise and fall pattern but then settles
into power law growth with an exponent of 8.7.}
\end{figure}

An interesting mixed case is found in the data for Belgium. It starts
out as typical rise and fall pattern but then, due to stricter lockdown
measures settles into a power law growth. This behavior is shown in
Fig. \ref{fig:Local-slope-Belgium}. Perhaps the rise and fall pattern
which we have discovered for various countries will eventually follow
the same behavior as Belgium or perhaps the rise and fall pattern
does move straight to the saturation. Currently it is too early to
confirm either hypothesis.

\begin{figure}
\includegraphics[scale=0.6]{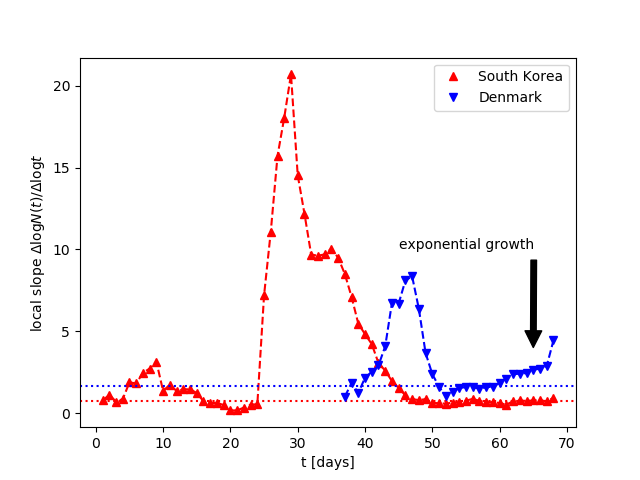}

\caption{\label{fig:Local-slope-exceptions}Local slope for countries exhibiting
exceptional behaviors. South Korea managed to ward off an exponential
growth and keeps the rate of infection at a power law growth of 0.72
(sub linear growth). On the opposite behavior: Denmark which at first
managed to curb the exponential growth to a power law has now started
to exhibit exponential growth again.}
\end{figure}

In Fig. \ref{fig:Local-slope-exceptions} two countries showing exceptions
to the patterns found in other countries are shown. First, South Korea,
where through wide range testing and strict isolation of infected
individuals the government has managed to reduce the explosive exponential
growth down to a power low growth with an exceptionally low exponent
of 0.72. And on the other hand Denmark, where early governmental measures
proved effective and the exponential growth was reduced to a power
law of 1.64 for a while only to start to grow exponentially again
soon thereafter.

\begin{figure}
\includegraphics[scale=0.6]{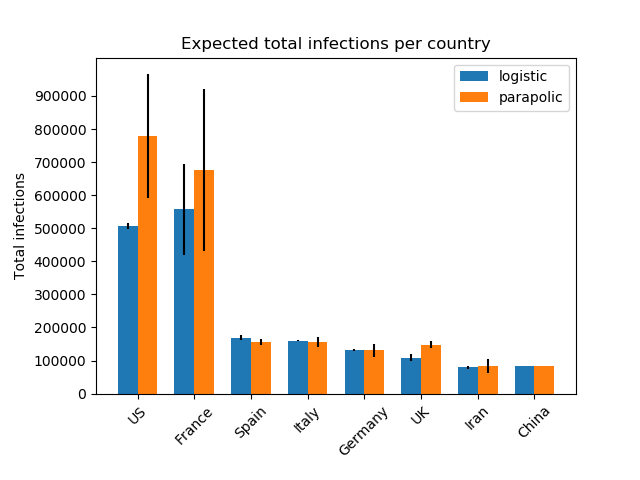}

\caption{\label{fig:Expected-final-numberLarge}Expected final number of total
infections per country for countries with a large number of infections.}
\end{figure}

\begin{figure}
\includegraphics[scale=0.6]{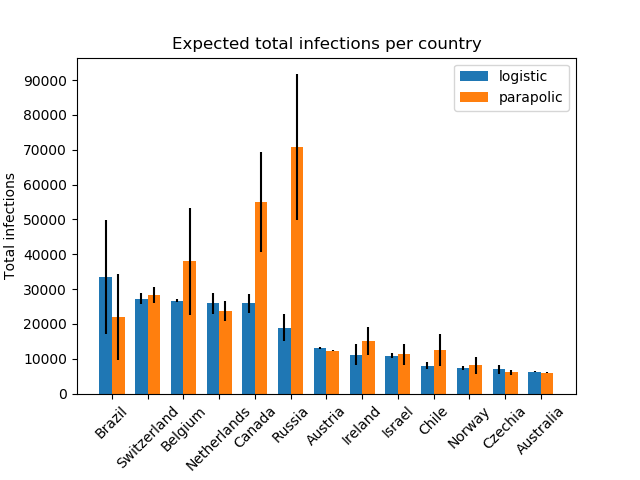}

\caption{\label{fig:Expected-final-numberSmall}Expected final number of total
infections per country for countries with a smaller number of infections.}
\end{figure}

The extrapolated final number of infections for both extrapolations
methods are shown in Fig. \ref{fig:Expected-final-numberLarge} for
highly affected countries and Fig. \ref{fig:Expected-final-numberSmall}
for lesser affected countries.

As can be seen the extrapolations generally predict the same order
of magnitude apart from the US, France, Canada, and Russia. The variability
of the extrapolated values depending on the considered number of points
is larger for the parabolic extrapolation. In the majority of the
considered data sets the parabolic extrapolations calculates larger
final numbers of infections compared to the logistic fit.

\begin{figure}
\includegraphics[scale=0.6]{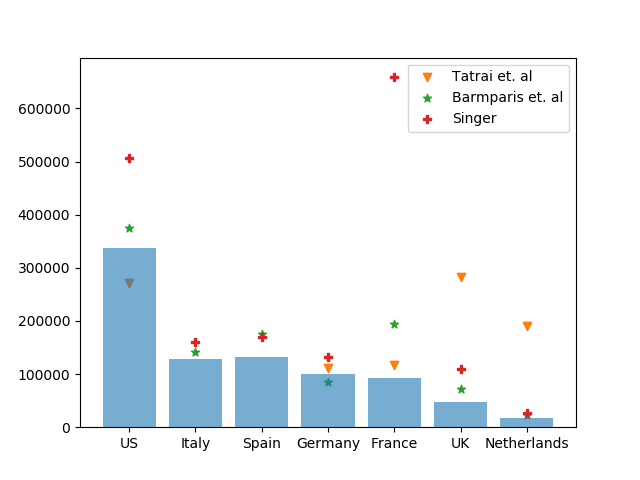}

\caption{\label{fig:ExtrapolationComparison}Current total number of infections
and extrapolations from different research groups. Only countries
where predictions were available for \cite{ttrai2020covid19,barmparis2020estimating}
are displayed. The spread in predicted final values is considerable. }
\end{figure}

In Fig. \ref{fig:ExtrapolationComparison} the current number of infections
is plotted with the extrapolated estimates of different studies \cite{ttrai2020covid19,barmparis2020estimating}.
As can be seen the predicted final number of infections varies considerably.
Generally, though, the extrapolations in this study seem closer to
the values of \cite{barmparis2020estimating} than to \cite{ttrai2020covid19}.
Particularly for the UK and the Netherlands the values reported \cite{ttrai2020covid19}
deviate strongly from our calculations as well as those reported in
\cite{barmparis2020estimating}. One outlier is our extrapolation
for France. While calculations with earlier data have produced numbers
in the order of 200'000-300'000 in line with the estimates of \cite{ttrai2020covid19,barmparis2020estimating}
the latest data points have drastically changed the extrapolation.
Whether this is a fluke or, indeed, a solid trend remains to be seen.
In the case of the US the values reported in \cite{ttrai2020covid19}
are much lower than the one extrapolated in our study and the calculated
values in \cite{barmparis2020estimating} and are already overtaken
by the current number of infections. 

\begin{table*}
\caption{\label{tab:Extrapolated-final-number}Extrapolated final number of
total infections $\overline{N}_{\infty}$ and predicted date $\overline{t}$
relative to the first day of infection for the logistic fit in that
particular country. Note that $\overline{t}$ and the corresponding
expected final date denotes the time when 95\% of the saturation value
has been reached and not when no new cases will be reported. Additionally
the local slope shape is indicated and the final percentage of infected
individuals of the logistic fit relative to the population of the
country. Countries exhibiting a power law growth have their scaling
exponent $\gamma$ indicated as well.}
\begin{tabular}{|l|c|c|c|c|c|c|c|c|} 
\hline Country & $\overline{N}_{\infty,\text{log}}$& $\overline{N}_{\infty,\text{par}}$&$\overline{t}_{\text{log}}$ &$\overline{t}_{\text{par}}$ &expect. date (log)&shape&$\gamma $& \% infected\\ \hline \hline 
US & $ 506629 \pm 9473 $ & $ 779575 \pm 188109 $ & $ 84 \pm 1 $ & $ 93 \pm 5 $ & 2020-04-15$  \pm 1 $ & exp &  & 0.04\\\hline
France & $ 557149 \pm 137282 $ & $ 676753 \pm 244622 $ & $ 107 \pm 3 $ & $ 136 \pm 12 $ & 2020-05-10$  \pm 3 $ & exp &  & 0.85\\\hline
Spain & $ 169256 \pm 6601 $ & $ 155401 \pm 9557 $ & $ 74 \pm 1 $ & $ 73 \pm 2 $ & 2020-04-15$  \pm 1 $ & power law & 6.18 & 0.36\\\hline
Italy & $ 159997 \pm 2298 $ & $ 155650 \pm 15781.77 $ & $ 78 \pm 1 $ & $ 80 \pm 6 $ & 2020-04-18$  \pm 1 $ & exp &  & 0.26\\\hline
Germany & $ 131912 \pm 1950 $ & $ 130537 \pm 20419 $ & $ 80 \pm 0 $ & $ 81 \pm 5 $ & 2020-04-16$  \pm 0 $ & power law & 11.45 & 0.16\\\hline
United Kingdom & $ 109012 \pm 10847 $ & $ 148293 \pm 10178 $ & $ 84 \pm 2 $ & $ 92 \pm 2 $ & 2020-04-24$  \pm 2 $ & power law & 9.69 & 0.16\\\hline
Iran & $ 80270 \pm 4580 $ & $ 84316 \pm 21460 $ & $ 60 \pm 2 $ & $ 66 \pm 12 $ & 2020-04-19$  \pm 2 $ & power law & 2.95 & 0.1\\\hline
China & $ 83108 \pm 143 $ & $ 82865 \pm 150 $ & $ 76 \pm 3 $ & $ 78 \pm 4 $ & 2020-04-07$  \pm 3 $ & power law & 2.48 & 0.01\\\hline
Belgium & $ 26710 \pm 538 $ & $ 37923 \pm 15434 $ & $ 72 \pm 1 $ & $ 79 \pm 10 $ & 2020-04-16$  \pm 1 $ & exp/power law &  & 0.23\\\hline
Russia & $ 18968 \pm 3916 $ & $ 70878 \pm 60969 $ & $ 84 \pm 2 $ & $ 105 \pm 18 $ & 2020-04-24$  \pm 2 $ & power law & 11.04 & 0.01\\\hline
Brazil & $ 33446 \pm 16351 $ & $ 21991 \pm 12412 $ & $ 58 \pm 9 $ & $ 54 \pm 13 $ & 2020-04-24$  \pm 9 $ & exp/power law &  & 0.02\\\hline
Switzerland & $ 27284 \pm 1693 $ & $ 28325 \pm 2275 $ & $ 53 \pm 3 $ & $ 60 \pm 5 $ & 2020-04-18$  \pm 3 $ & logistic &  & 0.32\\\hline
Netherlands & $ 25942 \pm 3054 $ & $ 23712 \pm 2916 $ & $ 54 \pm 4 $ & $ 53 \pm 5 $ & 2020-04-21$  \pm 4 $ & power law & 3.51 & 0.15\\\hline
Canada & $ 25896 \pm 2740 $ & $ 55110 \pm 14337 $ & $ 84 \pm 1 $ & $ 100 \pm 7 $ & 2020-04-18$  \pm 1 $ & exp &  & 0.15\\\hline
Austria & $ 13063 \pm 188 $ & $ 12296 \pm 140 $ & $ 44 \pm 1 $ & $ 44 \pm 1 $ & 2020-04-09$  \pm 1 $ & logistic &  & 0.15\\\hline
Israel & $ 10877 \pm 667 $ & $ 11244 \pm 3060 $ & $ 52 \pm 2 $ & $ 54 \pm 7 $ & 2020-04-13$  \pm 2 $ & logistic &  & 0.13\\\hline
Australia & $ 6213 \pm 149 $ & $ 6022 \pm 154 $ & $ 74 \pm 1 $ & $ 75 \pm 2 $ & 2020-04-09$  \pm 1 $ & exp &  & 0.02\\\hline
Norway & $ 7349 \pm 600 $ & $ 8107 \pm 2518 $ & $ 53 \pm 5 $ & $ 69 \pm 29 $ & 2020-04-19$  \pm 5 $ & power law & 2.23 & 0.14\\\hline
Ireland & $ 11197 \pm 2905 $ & $ 15096 \pm 3968 $ & $ 59 \pm 6 $ & $ 85 \pm 14 $ & 2020-04-28$  \pm 6 $ & logistic &  & 0.23\\\hline
Chile & $ 8025 \pm 971 $ & $ 12530 \pm 4611 $ & $ 50 \pm 3 $ & $ 71 \pm 15 $ & 2020-04-22$  \pm 3 $ & power law & 4.25 & 0.04\\\hline
Czech Republic & $ 6940 \pm 1214 $ & $ 6085 \pm 612 $ & $ 50 \pm 6 $ & $ 49 \pm 4 $ & 2020-04-20$  \pm 6 $ & power law & 3.54 & 0.06\\\hline 
\end{tabular}
\end{table*}

The numerical results of both extrapolation methods including the
estimated date of saturation and the estimated percentage of people
infected relative to the country's population is given in Table \ref{tab:Extrapolated-final-number}.
Additionally we have indicated the growth pattern (shape) of the analyzed
data set and the scaling exponent where the local slope analysis showed
a power law growth. It should be noted that the final date indicated
corresponds to the time when 95\% of the saturation value has been
reached and not when no new cases will be reported (which in the case
of the logistic curve happens only for $t\rightarrow\infty$).

It is to be assumed, given the behavior of Belgium in Fig. \ref{fig:Local-slope-Belgium}
that predictions made for countries showing a rise and fall patterns
are not accurate. The slowdown might be just part of the transition
to a slower power law growth. Countries, however, that show a good
logistic fit such as Switzerland, Austria and others (see Table \ref{tab:Extrapolated-final-number})
might prove to have quite an accurate prediction. Expected saturation
numbers for countries showing a power law growth and subsequent slowdown
would likely be less accurate than countries with logistic patterns
but still more accurate and trustworthy than countries with a rise
and fall pattern.

\begin{figure}
\includegraphics[scale=0.6]{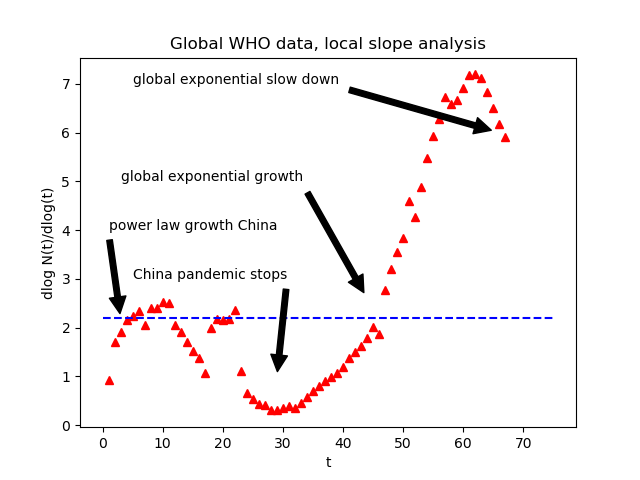}

\caption{\label{fig:Local-slope-analysisWorld}Local slope analysis of the
global WHO data for the total number of infections. The local slope
offers a microscopic and amplified view of the events of the global
pandemic. The last few data points indicate a slow down of the exponential
growth.}
\end{figure}

The same local slope analysis can be performed for the aggregated
world wide total number of infections reported by the WHO in Fig.
\ref{fig:Local-slope-analysisWorld}. As can bee seen the history
of the disease spread is accurately recovered form the local slope
plot. The beginning is marked by a power law growth in mainland China
with the subsequent saturation and stop of further growth. At this
point the number of infected people world wide picks up and the rate
of infections is shown to grow strongly exponentially. The last few
data points, however reveal a reversal of the growth trend to an exponential
slow down. Currently 3 of the top 5 most affected countries (USA,
France and Italy) show such as rise and fall exponential behavior
and only Spain and Germany have shown a power law growth. Given that
the USA has the strongest growth rate and highest number of infections
totaling currently at about 25\% of the world wide total number it
is not surprising to see that the world data follows essentially the
same behavior as the USA.

While, indeed, the last few data points indicate that the strong exponential
growth is slowing down it cannot be predicted whether the subsequent
period will stabilize in an power law growth or a rapid decay towards
saturation for the world data as well as for the countries exhibiting
the rise and fall growth pattern such as USA, France, and Italy.

\section*{Discussion \& Conclusions}

We have analyzed the top 25 most affected countries with the local
slope analysis of the total number of infections and have found that
different countries exhibit substantially different growth patterns.
We have identified three different patterns: exponential rise and
fall, power law, and logistic. For the countries following a power
law growth we have determined the country-specific scaling exponent
and summarized in Tables \ref{tab:Countries-exponents} and \ref{tab:Extrapolated-final-number}.

For countries that exhibit a clear slowdown in the rate at which the
infections spreads we have extrapolated a tentative final number of
infections by two different extrapolations methods: logistic and parabolic.
While it has been shown that only few countries such as Austria and
Switzerland actually follow a logistic growth behavior the extrapolation
with the logistic curve was still justified for other countries since
the fit closer to the saturation seems to become increasingly better
\cite{jia2020prediction,yang2020rational,wu2020generalized,batista2020.02.16.20023606}.
We have used these extrapolation methods purely on numerical grounds
and do not ascribe them the ability to actually explain the physical
and biological reasons why the spread would saturate exactly at those
given points. With few exceptions both extrapolations methods have
calculated the same order of magnitude of final numbers of infections
and also agree on the final date when this point will be reached (95\%
of $N_{\infty}$).

It should be noted that while we have selected only countries that
showed a significant slowdown in the local slope plot this does not
mean that the data really moves to a saturation. It might simply mean
that the further growth will settle to a power law growth once the
governmental measures and lockdowns have started to take effect. A
slow down, thus, can mean that the growth has now been reduced to
a power law when most if not all random encounters with other people
have been eliminated. Only time will tell if this state of lockdown
can be maintained long enough that a real saturation of total infections
can be reached or if a second exponential wave will start upon relaxing
the lockdown measures too early.

It can also simply mean that the data currently shows a downward trend
just to pick up speed again a few days later as can be seen in the
data for the USA at around day 50 (Fig. \ref{fig:Local-slope-plotUS})
and in data from Belgium at around day 33 (Fig. \ref{fig:Local-slope-Belgium}). 

So far most extrapolations and predictions on the true extent of the
global pandemic (excluding China) for example \cite{barmparis2020estimating,ttrai2020covid19}
seem to have severely underestimated the final numbers of infections.
So much so that the extrapolations of final numbers of infections
of only a few days ago are already overtaken by the current number
of infections. Also it has been found that the logistic fit generally
seems to underestimate the final number of infections \cite{yang2020rational}.
We do not expect the extrapolations in this study to fare any better.
We believe, however, that the predictions for the countries that actually
exhibit a logistic pattern such as Austria and Switzerland to be relatively
accurate. Much less accurate but likely to be in the right order of
magnitude are the countries exhibiting a power law pattern with a
recent slowdown. The least accurate predictions and probably even
underestimating the order of magnitude are the countries following
an exponential rise and fall pattern.

Why then are the extrapolations so wrong? Firstly the calculations
of the saturations are based on the fitting of some nonlinear function
that determines its parameters based on the previous data points.
As we have shown however there is hardly any evidence that for example
the logistic model in eq. (\ref{eq:logeq}) describes the actual progression
of the pandemic (with a few notable exceptions). Secondly, the physical
and biological limit of possible infections is not reached by far.
Current population infections are in the order of 0.2\% of the total
population and are expected by the extrapolations to stay below 1\%
(Table \ref{tab:Extrapolated-final-number}). That is far too little
to effectively limit the spread by herd immunization or actual lack
of susceptible individuals. Thirdly, most models do not take into
account lockdown measures and interruption of disease spreading paths
as instituted in many countries and therefore predict with the first
slow down the beginning of the end of the pandemic. As shown in Fig.
\ref{fig:Local-slope-Belgium} for Belgium this is simply not true.
Finally, one should not forget that power law growth, while not exponential,
is still growth. And it will still take a long time to reach the point
where an actual saturation has been achieved. So far, outside mainland
China, the only country that has effectively managed to control the
spread of the disease is South Korea. But even there the number of
infections follows a power law with an extraordinarily low exponent
of 0.7 (sub-linear growth) and no further slowdown seems to be in
sight. Also it should be noted that even in China the pandemic is
not over, it is just managed: every day China registers between 50
and 100 new cases. Reaching the saturation does not mean the end of
the pandemic. Leaving those few remaining daily new cases uncontrolled
and without strict isolation management, or easing the strict measures
too early can lead to a second, perhaps more devastating wave. A cautionary
example is Denmark that managed to reduce the initial exponential
growth to a power law growth of 1.64 only to grow again exponentially
soon thereafter (Fig. \ref{fig:Local-slope-exceptions}).

Therefore the predictions made should really only be taken as current
extrapolations in the best case scenario that the current slowdown
is actually an indication of a coming saturation and not just noise
or transitions into a different growth patterns. The most likely outcome,
however, is that the countries showing a rise and fall pattern are
simply on the way to adjusting to a slower spread with a power law
as has been shown in the case of Belgium (Fig. \ref{fig:Local-slope-Belgium})
but will not reach the saturation range any time soon.

Finally the analysis of the world wide spread of the infection shows
that the trend of a slowdown is a reflection of the the biggest contributor,
the USA currently contributing 25\% of all cases. As long as the US
is the largest contributor the world data is likely to behave in a
similar way as the infections reported in the USA. One should, however,
not neglect that the numbers of infections given in the world data
are much less accurate than those reported by single countries. In
particular countries with few medical facilities or the lacking ability
to test for infections of the COVID-19 virus are bound to severely
under-report the true extent of the pandemic. Therefore it can only
be assumed that the actual number of infections is much higher and
while the trend for individual countries, particularly in Europe and
the USA might point towards a saturation many countries on other continents
might not have the means to manage the spread of the disease. Due
to the current low percentage of the infections relative to the population
this might initiate a second wave of a global pandemic as soon as
the strict lockdown measures are relaxed and travels into countries
with less accurate reporting are allowed again.
\begin{acknowledgments}
The author reports no funding related to this research and has no
conflicting financial interests.
\end{acknowledgments}

\bibliographystyle{unsrtnat}
\bibliography{covid01}

\end{document}